\documentclass[]{interact}

\usepackage[natbibapa,nodoi]{apacite}
\theoremstyle{plain}% Theorem-like structures provided by amsthm.sty

\theoremstyle{definition}

\theoremstyle{remark}

\begin{document}
	
\title{A high-performance MEMRISTOR-based Smith-Waterman DNA sequence alignment Using FPNI structure}
	
\author{
\name{Mahdi Taheri\textsuperscript{a}\thanks{\textsuperscript{a} M. Taheri. Email: mahditaheri@eng.uk.ac.ir} , Hamed Zandevakili\textsuperscript{b}\thanks{\textsuperscript{b} H. Zandevakili. Email: H.Zandevakili@eng.uk.ac.ir} . Ali Mahani\textsuperscript{c}\thanks{\textsuperscript{c} A. Mahani. Email:  amahani@uk.ac.ir} }
\affil{Department of Electrical Engineering, Shahid Bahonar University, Kerman, Iran}
}

	\maketitle
	
	\begin{abstract}
	\textbf{Purpose-}

	This paper aims to present a new re-configuration sequencing method for difference of read lengths that may take place as input data in which is crucial drawbacks lay impact on DNA sequencing methods.
	\newline
	\textbf{Study design/methodology/approach-}
	
	We propose a new Race-logic implementation of the seed extension kernel of BWA-MEM alignment algorithm. It is the first proposed method that does not need re-configuration to execute the seed extension kernel for different read lengths as input. We use MEMRISTORs instead of the conventional complementary metal–oxide–semiconductor (CMOS) which leads to lower area overhead and power consumption. Also, we benefit from Field-Programmable Nanowire Interconnect Architecture as our matrix output resulting in a flexible output which bypasses the reconfiguration procedure of the system for reads with different lengths. 
	\newline
	\textbf{Findings-}
	
	With considering the power, area and delay efficiency, we gain better results in comparison with other state-of-the-art implementations. Consequently, we gain up to 22x speed-up in comparison with the state-of-the-art systolic arrays, 600x speed up considering different seed length of the previous state-of-the-art proposed methods, at least 10x improvements in area overhead, and also, {$10^5$}x improvements in power.
	\newline
	\textbf{Originality/value-}
	
	A new memristor based smith-waterman matrix implementation is proposed in this work. We shows our design give this flexibility to get the matrix output depends on the different input dimension without suffering from the extra latency.

	\end{abstract}
	
	\begin{keywords}
		Bioinformatics, BWA-MEM, memristor, FPNI, Race logic. 
	\end{keywords}
	
	\section{Introduction}
	Based on the recent researches on the genomic sequence
	alignment, there are variety of algorithms and specific
	designs to make better performance and energy consumption of the sequencing aligners.
	We can put genomics in the group of big data science and by growing technology, it is getting much bigger.
	The volume of produced data by genomics can be compared with three main Big data generators\citep{giles2012computational}:
	\begin{enumerate}
		\item astronomy: Over these decades, Astronomy is placed in the group of Big Data challenging.
		\item YouTube: There is huge number of sharing stuffs and videos that are created and shared in YouTube
		\item Twitter: Makes a lot of opportunities for new insights by mining more than 400 million messages that are sent every day  .
	\end{enumerate}
	
\begin{figure}[h]
		\includegraphics[width=1\columnwidth]{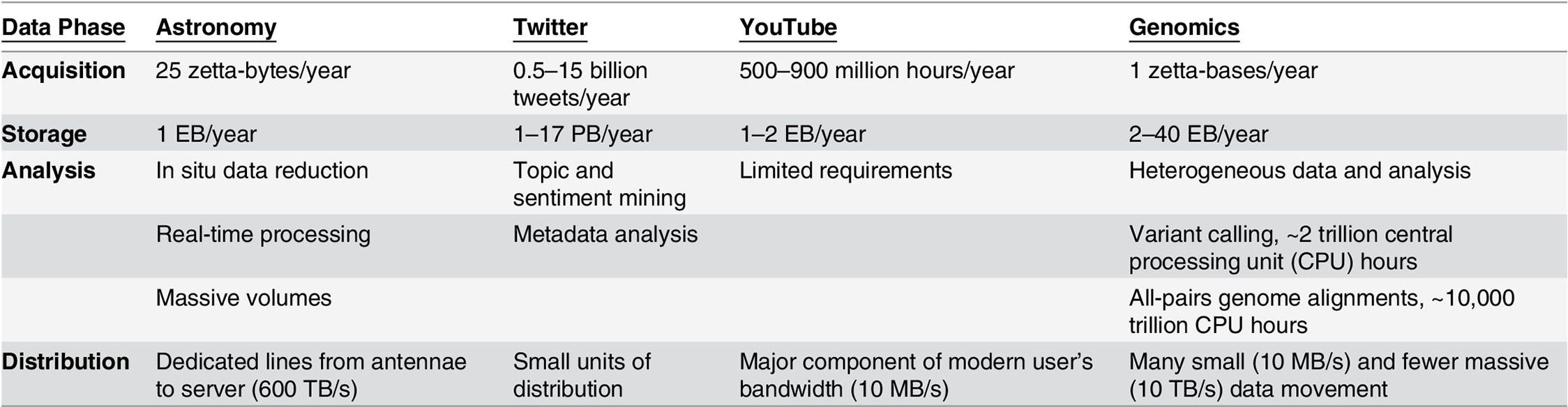}

		\caption{Comparison of Four groups of Big Data in 2025 are shown in this Figure
			\citep{stephens2015big}.}
		\label{Fig. figures/F2.PNG}
	\end{figure}

	In Fig. \ref{Fig. figures/F2.PNG} \citep{stephens2015big} a comparison of these four group of data generators is given that shows how genomics is increasingly overcomes in case of demanding data acquisition, storage, distribution, and analysis.
	
%	The amount of data that is produced by genomics is growing massively based on the previous discussion, and resulting genomics can be in order of bench of gigabytes. Thus the computation volume of data requires increasingly long time calculation. Therefor, genomics data analysis is becoming a large Data bottleneck. Accordingly, acceleration of genomics data algorithms is an important problem.

	The first step for most of the genomics applications is sequence alignment.
	there is lots of reads of DNA strand which have to be aligned against reference genome and the best alignment for each read is produced as output. There are variety of sequence alignment tools such as :
	
	\begin{enumerate}
		\item Bowtie \citep{langmead2009ultrafast}
		\item BWA \citep{li2009fast}
		\item MAQ \citep{li2008mapping}
		\item SOAP \citep{li2009soap2}
		\item BWA-MEM \citep {li2013aligning}
	\end{enumerate}
	
	Consider the state that we want to find all local alignment by using dynamic programming approach as an example of the alignment algorithms. If we choose Smith and Waterman algorithm \citep{smith1981identification}, which uses O(nm) time for aligning a read of length n against a reference of length m, it can be concluded that the approach is too slow.
	
	For example, NGS as the fastest sequencing, takes about hours with a lot of memory usage to sequence an entire human DNA.
	Based on the experimental results of the \citep{lam2008compressed}, aligning 1000 characters as a read against human genome will takes more than 15 hours.

	In case of real application, we works with genes or chromosomes that are about few thousands to a few hundred million length. If we align the hole human genome with SW method, it will lasts for about days to weeks.
	
	There are other algorithms like BLAST \citep{kent2002blat} which are heuristic methods.
	They are used to find local alignments very efficient.
	
	In case of using BLAST, it takes 10-20 seconds to align a read of 1000 bp against the human genoem \citep{kent2002blat}.
	
	It is obvious that with this 
	Time-consuming calculations, general purpose processors are not good solution for executing these bioinformatics workloads.
	Thus we need more parallel and specific hardware such as GPU or FPGA that are dedicated to massively accelerate the intensive computations and lead to large speedups.
	
	In this work, We accelerate the Smith-waterman-like algorithm with race logic strategy based on memristor elements to achieve speedup of the execution time.
	The rest of this paper is organized as follows:We provides related works in section II. Our  design contributions and details of our MEMRISTOR-based design are discussed in Section III. Section IV evaluates  the  results  and  finally,  Section  V concludes this article.
	
	\section{Related work}
	
	We are experiencing an exponential growth of experimental data and information in Biology, which is called data explosion\citep{marx2013biology}. One of the most useful operations in Bioinformatics is DNA sequencing. There are four nucleotides (A, C, G, T) that make the foundation of the DNA sequences. swapping these nucleotides, cause alternate biochemical functions and products within the DNA. One of the most Severe computational part of Bioinformatics is to find similarities between two DNA sequences which is called pairwise alignment.
	There are different methods accomplish this for Biologists which leads to different time consumption. The Smith-Waterman (SW) is one of the most accurate algorithms with high sensitivity degree but high computational time and high hardware resource usage. Consider that the complexity of SW is of quadratic order.
	The BLAST\citep{altschul1997gapped} and FASTA\citep{pearson1988improved} are derivative methods of SW which do not lead to optimal solutions because of sensitivity loss but significantly faster.
	Another dynamic programming method for comparing two macro molecules is Needleman-Wunsch algorithm (NW)\citep{wilbur1984context} which calculate the alignment score between two sequence based on the Levenshtein distance.
	There is different other efforts to reduce computational time of different parts of the pairwise alignment algorithms.
	A custom ASIC implementation of a BioSCAN is introduced in \citep{singh1996bioscan} in which  heuristic and very high density implementation caused in high performance. A new method of information representation was proposed in \citep{madhavan2014race} that performs computation by setting up logical race conditions in a circuit on ASIC platform and they achieved about 3x higher throughput at 5x lower power density.
	The authors in \citep{rucci2018swifold} evaluates SWIFOLD:a Smith-Waterman parallel Implementation for Long DNA sequences which is implemented on Intel core with OpenCL and they claim that their method increases better performance with higher resource consumption. In another work, In \citep{zokaee2018aligner} a ReRAM-based process-in-memory architecture is designed to improve short read alignment throughput per Watt by $13\times$. The authors in \citep{alser2017gatekeeper} propose a new hardware accelerator in which the most incorrect candidate locations fills out with 130-fold speedup than software.There is a faster implementation of SW in \citep{farrar2006striped} which achieves $2 to 8\times$ performance improvement in comparison of other SIMD based SW implementations.Also, intrinsic delay of the circuits edit-distance computation elements as in \citep{banerjee2018asap} was utilized to propose the ASAP accelerator based on the RACE-logic hardware acceleration that is presented in \citep{madhavan2014race} for accelerating SW and NW algorithms on an ASIC platform.Their work leads to $200\times$ speedup than an equivalent Smith-Waterman-C implementation.There are some other works that Accelerated BWA-MEM genomic Mapping Algorithm on different platforms such as GPU and FPGA. BWA-MEM is widely used algorithm to map genomic sequences onto a reference genome. This algorithm is composed of three main computational kernel\citep{li2013aligning} :

	\begin{enumerate}
		\item SMEM Generation: This kernel is used to find seeds (sub-strings of the reads) that are likely mapping locations of the read against the reference genome. There is a chance of generating several seeds with variable length for each read~\citep{li2009fast}.  This step is an exact-match-finding phase that uses the Burrows-Wheeler transform. For this work, seeds are at least nineteen characters and a maximum of 131.
		
		\item Seed Extension: This steps is an inexact-matching step that executes chaining and extending of seeds in two directions by using a Smith-Waterman-like algorithm~\citep{smith1981identification}. This part of the BWA-MEM algorithm finds the optimal local alignment by using a scoring system. 
		
		\item Output Generation: In this step the best alignment (i.e., the one with the highest score) is finalized and provided as the output in SAM-format, if necessary.
	\end{enumerate}
	
	Note that the seed extension kernel used in BWA-MEM is different from the Smith-Waterman algorithm in two substantial ways~\citep{houtgast2015fpga}: (1) Non-zero initial values: The initial values in the first column and row depend on the alignment score of the seed found by the SMEM Generation kernel. (2) Additional Output Generation: Other than the local and global alignment scores, the exact location inside the similarity matrix and a maximum offset (indicating the distance from the diagonal at which a maximum score has been found) are also generated.
	
	Several techniques have been proposed to accelerate the BWA-MEM inexact alignment algorithm. However, the seed extension step of this algorithm makes it inherently a slow design\ref{tbl: BWA}.
	
	\begin{table}[h]
%		\captionsetup{justification=centering}
		\caption{Profiling the BWA-MEM algorithm \citep{houtgast2015fpga}.}
		\label{tbl: BWA}
		\centering
		\resizebox{0.5\columnwidth}{!}{
			\begin{tabular}{c c c c}
				\textbf{Kernel} & \textbf{\% Execution time} & \textbf{Bound}  \\
				\hline
				SMEM generation  & 56\% & \textbf{Memory}\\
				Seed extension & 32\% & \textbf{Computational} \\
				Output generation & 9\% & \textbf{Memory}\\
				Other & 3\% & \textbf{I/O}\\
				\hline
		\end{tabular}}
	\end{table}
	
	The first accelerated implementation of BWA-MEM is presented in \citep{houtgast2015fpga} with evaluating a number of FPGA-based systolic array architecture. Their implementation is $3\times$ faster than the software-only execution.
	A hardware acceleration of the BWA-MEM genomics short read mapping for longer read length is implemented in this article~\citep{houtgast2018hardware}. This design is based on a previously proposed architecture ~\citep{houtgast2015fpga}, where an FPGA-based 1D-systolic array is used to accelerate the BWA-MEM genomics mapping algorithm. The main idea is to insert some exit points between the PEs of the 1D-systolic array to avoid unnecessary calculations for shorter reads. By doing so, shorter reads do not have to go through all of the PEs and can exit the array once they get to the first exit point.
	The authors discussed acceleration of the Seed Extension kernel of the BWA-MEM algorithm on a GPU accelerator and achieved up to $1.6\times$ improvement in comparison of application-level execution time\citep{houtgast2016gpu}.
	Power-Efficiency Analysis of Accelerated BWA-MEM Implementations on Heterogeneous Computing Platform against the software-only baseline system is studied in \citep{houtgast2016power} By offloading the Seed Extension phase on an accelerator. A high-performance FPGA-based Seed Extension IP core is designed\citep{phamhigh} for BWA-MEM DNA Alignment that achieve $350\times$ speedup when compare to an Intel Core i5 general purpose processor. 
	Authors gain up to $14.5\times$ speedup than the Smith-Waterman algorithm by :(a) Applying heuristics ; (b) Processing MEMs ; and, (c) Extracting MEMs by using a bit-level parallel method\citep{bayat2019pairwise}.
	This is considerable that after all these works, 
    The problem of memory accessory, area overhead, time and power consumption of the alignment algorithms methods and implementations are still extremely problematic.
    Thus, we aimed these problems in our work and by our suggested methods, we improved all of the above mentioned problems.
	\section{Proposed design}
	This section describes the proposed method for filling the similarity matrix of the Smith-waterman-based algorithm and shows how it can speedup the time and reduce power consumption in comparison of the state-of-the-art architectures.
	Besides, our method uses a unfixed length strategy that can leads to higher speedup due to it doesn't need to be reconfigured for different reads lengths.
	
	There is a new data representation that is used for broad class of optimization problem which is called "Race Logic". This method can be used for the kind of problems that use dynamic programming algorithms to be solved. There are different implementation of Race Logic such as synchronous and asynchronous which we focus on synchronous type for our design. Race Logic idea is based on the race conditions in a circuit to optimize computation in case of time. We designed a SW similarity matrix with the idea of the Race Logic design. Also we use MEMRISTOR instead of the conventional complementary metal–oxide–semiconductor (CMOS) which leads to better performance. In addition, we considered Field-Programmable Nanowire Interconnect\citep{zandevakili2018new} Architecture as our matrix output. 
	This is significant that we achieve to lower power consumption and area overhead due to using a MEMRISTOR structure in comparison of the previous CMOS, ASIC and FPGA structures that is mentioned in results. and also we gain lower delay As a result of 
		\begin{enumerate}
		\item Using MEMRISTOR structure
		that is using RACE logic strategy which leads to lower circuit delay 
		\item Utilization of FPNI as a flexible output which bypasses the reconfiguration procedure of the system for reads with different lengths.
	\end{enumerate}
	\subsection{Algorithm description}
	First we describe the main idea of our design and show how it can lead to proper answer of the Smith-waterman-like matrix with the performance improvement. As we know, Smith-Waterman algorithm is a dynamic programming algorithm that can compute the alignment score(Levenshtein distance) of two read and partial-reference genome string with the Q,R length respectively. For calculating the scoring alignment of these two strings, the algorithm construct a matrix S which is a lattice of size $I_{Q}\times I_{R}$ ($I_{Q}$, $I_{R}$ are the length of two strings) and with the recursive equation it can calculate the minimum edit distance between two strings. Notice that in BWA-MEM algorithm which is in our consideration for implementing our proposed design, the length of two strings are as same as each other and we have a Square matrix in each solution. But it's dimension may be different based on the reads length and we solve this problem by using FPNI as a flexible output of the circuit which help us to earn all the outputs of different matrix dimension without any problem to change the circuit of any re-configurations.

	\begin{equation}
	\label{eq: Min}
	\begin{array}{c} \vspace{8px}
	
	DP_{(i,i)} =MIN
	\begin{cases}
	DP_{(i-1,i-1)} +T_{(Match,Miss-match)}\\
	DP_{(i-1,i)} +T_{(Gap)} \\
	DP_{(i,i-1)} +T_{(Gap)}
	\end{cases}
	\end{array}
	\end{equation}
	where DP denotes the similarity matrix, $T_{(Match,Miss-match)}$ is the assigned score for when a match or a mismatch occurs (usually 0 for a match and a 2 for a mismatch \citep{banerjee2018asap}), and $T_{(Gap)}$ is the gap penalty with usual 1 value\citep{banerjee2018asap}.It worth mentioning that Match is for situation that two corresponding nucleotide are the same as each other and Miss-match is the state that they are not the same. Notice that we can choose these parameters in the way that optimize the accuracy of the alignment based on the structure of the sequences that are being compared \citep{wang2011comparison}\citep{henikoff1992amino}\citep{sung2009algorithms}. Besides, we use fixed penalties for the gap between nucleotides with the value that is more commonly used\citep{sung2009algorithms}.
	The above equation which is a representative of the Smith-Waterman similarity matrix local alignment leads to finding the largest sub-string of R which is mapped with string Q with the lowest Levenshtein distance(LD)(See \citep{navarro2001guided},\citep{levenshtein1966binary} for more information). Although this method is accurate and yield to optimal alignment with high computational complexity. To overcome this problem, we can replace the LD values in Equation\ref{eq: Min} with their equivalent  propagation delays and use the delay based approach for addition and minimization.
	Consider that, these two operations(addition and minimization) are necessary for recursive equation\ref{eq: Min}
	
	For more clearance, we give some examples of how the addition and minimization operations can be modeled by Race Logic strategy.
	\newline
	- Race logic:
	\newline
	Suppose that we have two signals(M and N) that are set to logic value '1' (inject a high signal) at different times.
	This time delay is representing the different values of these two signals.For example, consider that the signal M is set to '1' with a specific time delay(time delay = D1) that means the value of M is "D1" and the second signal is set after D2 second time delay(time delay = D2) that mean N value is "D2".
	\begin{enumerate}
		\item If we want to add these two values with each other, we can combine the circuit elements of M and N together in series. That mean the total propagation delay of the output is a result of adding "D1" with "D2".
		\item If we connect these two circuit elements to an OR gate, the signal that arrive first to OR gate, emerges out of that. This structure is a Viewer of minimization operator. Because both signals have the '1' value and the signal which have the less amount of delay, will arrives first to OR gate and make the output of this gate '1' earlier.
		\item For calculating the value of the output, we can place a counter at the end of our Race Logic design that serve as a decoder\citep{banerjee2018asap}.
	\end{enumerate}

		\begin{figure}[h]
		\centerline{\includegraphics[width=\columnwidth]{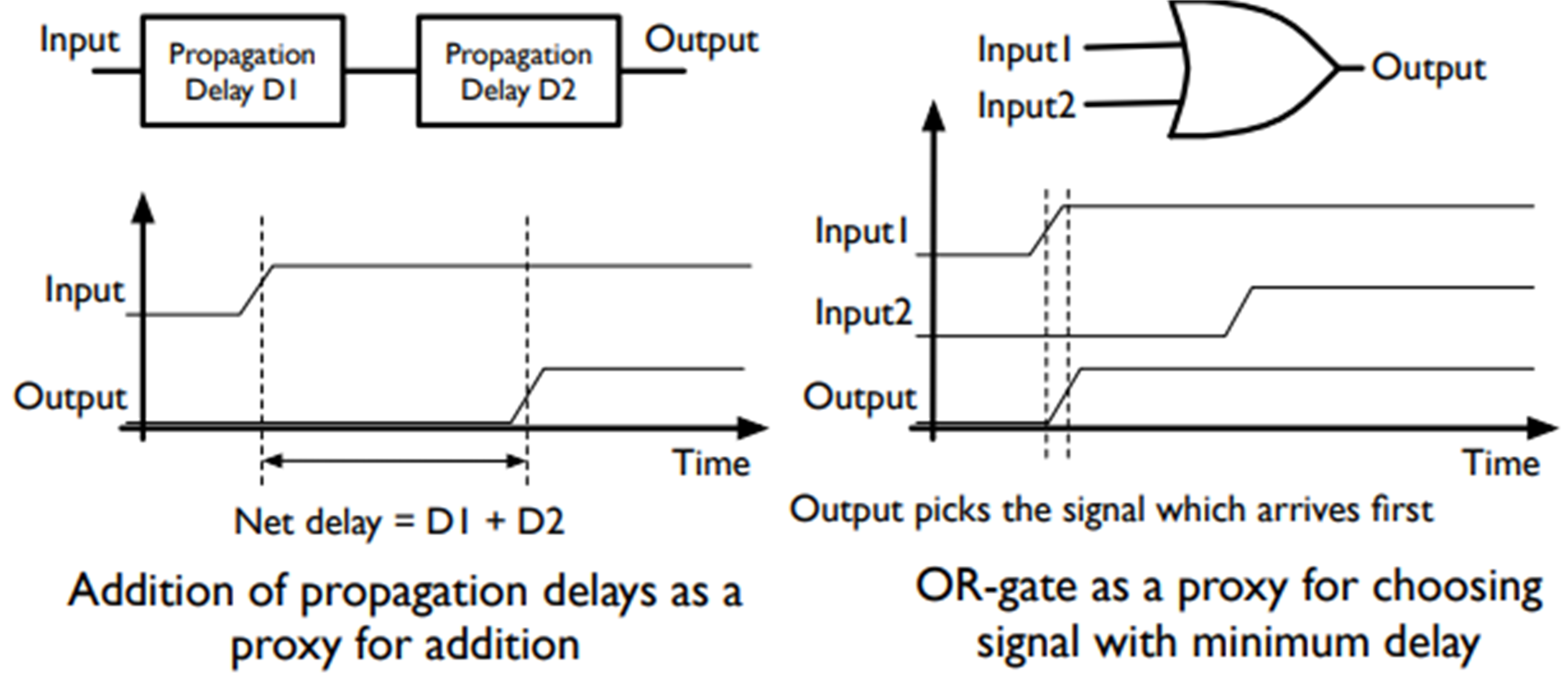}} 
%		\captionsetup{justification=centering}
		\caption{Computing with propagation delays: Delay-based proxy for the addition
        operator is a series connection, and the proxy for the min operator is the OR gate\citep{banerjee2018asap}.}
		\label{Fig. Race}
    	\end{figure}
	
	We can apply these delay-based computations to SW similarity matrix of LD calculation. So the delay between the rise edge of the input signal in the lattice and its emergence at any of the element on the last row that is the minimum score of the local alignment.
	%inserting some images of addition and minimization operations and a image of a hole matrix with Race Logic elements

	\subsection{Proposed architecture}

		\begin{figure}[h]
		\centerline{\includegraphics[width=\columnwidth]{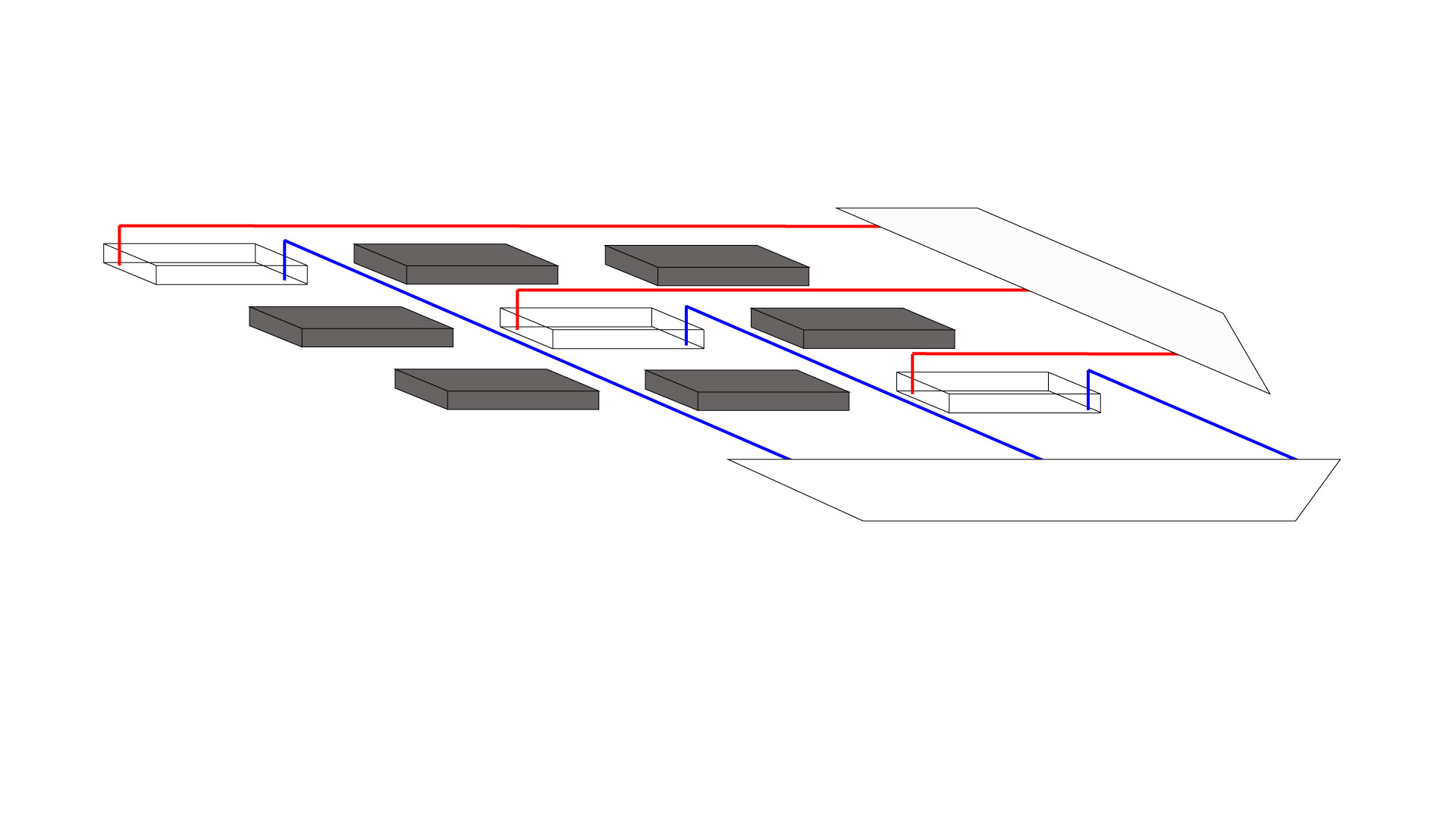}} 
%		\captionsetup{justification=centering}
		\caption{accelerated architecture.}
		\label{Fig. figures/swmatrix.JPG}
    	\end{figure}
	Fig \ref{Fig. figures/swmatrix.JPG} demonstrate our accelerated architecture. It includes some basic cells to implement the desired functionality, and a routing network to easy access to the output of some predefined basic cells. More details about the different parts of our proposed architecture will be presented in the following: 
    \newline
	•- MEMRISTOR-element:
	\newline
	Memristors \citep{chua1971memristor} are new two-terminal logical and scientific basis and fourth classical circuit elements as same as the resistor, inductor, and capacitor.
	
	Memristors are changeable resistors which can be used for memory. In this case, the resistance will stored as data. We can also use Memristive devices\citep{chua1976memristive} in other applications such as logic and analog circuits.

	We can refer to some points of using memristors instead of CMOS circuits in our Race Logic :

	\begin{enumerate}
		\item With these devices, we can read and write data faster than CMOS circuits\citep{torrezan2011sub}.
		\item They are typically small devices. Hence, the CMOS circuits usually bigger than the memristive-based circuits.
		\item Nonvolatility is the main feature of memristors and their compatible with standard CMOS technology\citep{borghetti2009hybrid}. They are eather ideal for FPGA-like applications
	\end{enumerate}
	
	From above, we can conclude that Memristive devices provides nonvolatile, dense, fast, and power efficiency to solving many major problems of the semiconductor devices.
	
\begin{figure}[h]
		\centerline{\includegraphics[width=\columnwidth]{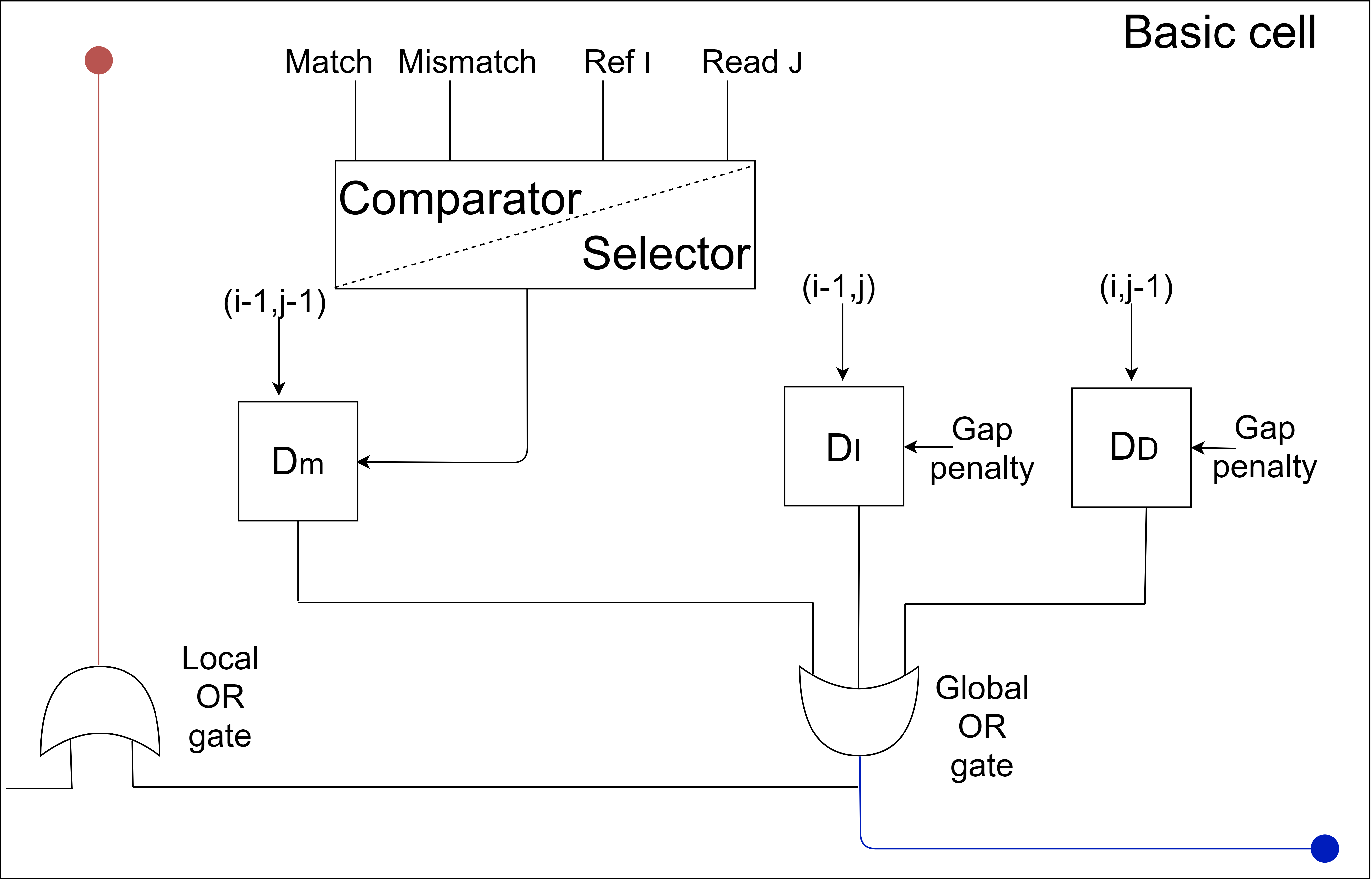}} 
%		\captionsetup{justification=centering}
		\caption{Basic cell of our proposed design.}
		\label{Fig. figures/Basic-cell.JPG}
    	\end{figure}

	Consider that we make a programmable design in which user can set the corresponding delay of "match", "mismatch", and "gap" penalties. For example, when we know that the most nucleotide comparisons are match, we can encode in the way that "match" delay has '0' time delay and this ensures us that large portions of our SW matrix is taken zero time to be explored.
	Different values for penalties help us to optimize the search time.
	\newline
	•- Basic cells:
	\newline
	The schematic of our proposed cell is shown in Fig. \ref{Fig. figures/Basic-cell.JPG}. Accordingly, it includes three delay elements (DM, DI, DD) which are responsible for the mathematical operations of the Eq. \ref{eq: Min}. respectively, a comparator/selector unit which has to compare the value of two nucloetides that are the inputs of each matrix cell and decides if match or mismatch occurs ,one local OR gate to implement the Min operation in Eq. \ref{eq: Min} , and one global OR gate to give us the flexibility of choosing output from different stages of the SW matrix.
	\newline
	
\begin{figure}[h]
		\centerline{\includegraphics[width=\columnwidth]{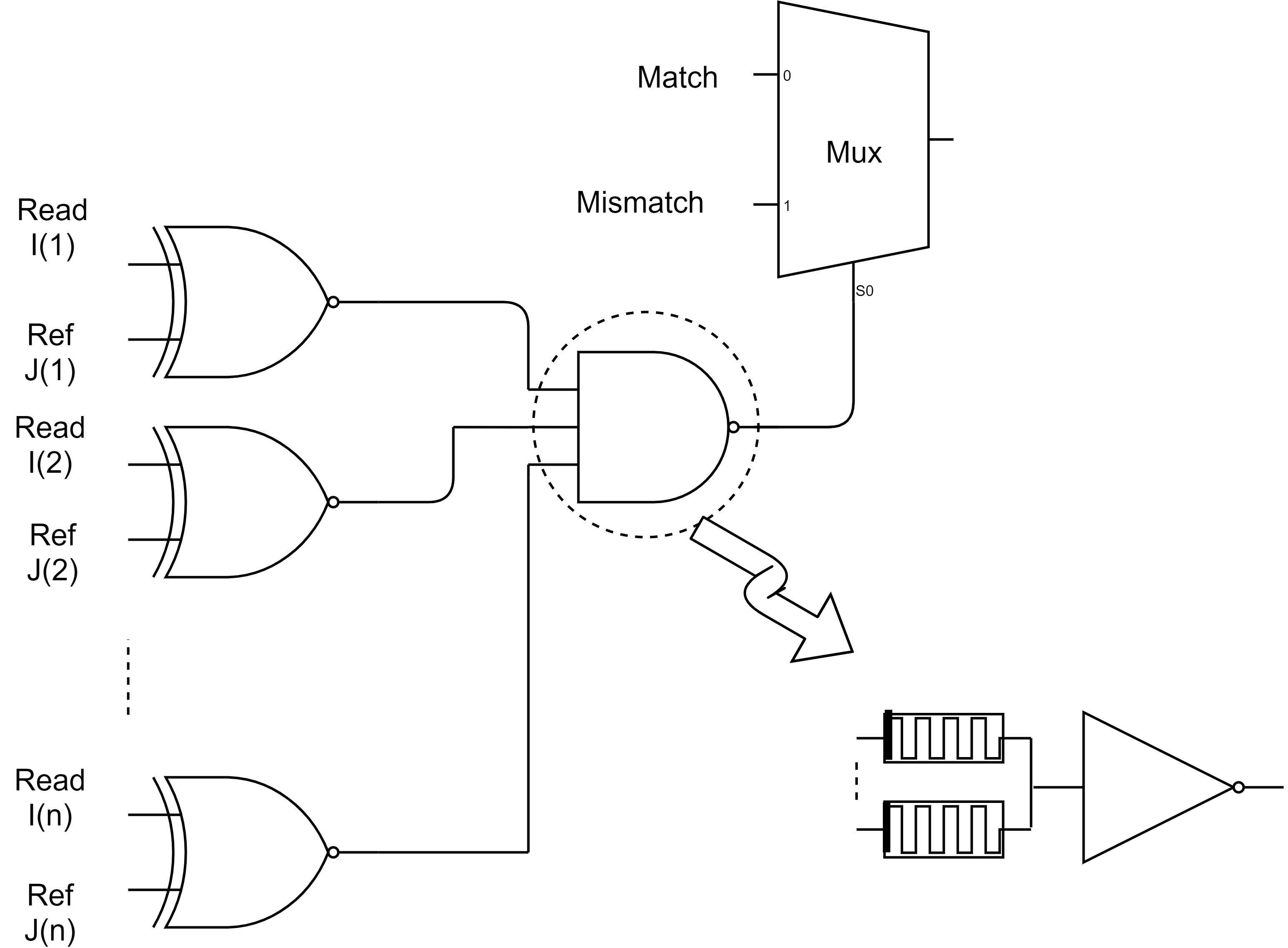}} 
%		\captionsetup{justification=centering}
		\caption{comparator/selector unit.}
		\label{Fig. figures/comperator.JPG}
     	\end{figure}

	•-	The comparator/selector unit:
	\newline
	This section includes several CMOS XNOR gates and a memristor-based NAND gate which are used to compare the “Ref” and Read” data Also the multiplexer controlled by the output of the comparator stage, that defines the corresponding match or mismatch penalty as its output. When the output value of the comparator becomes “0” this means the “Ref” data is equal to the “Read” data; and the proportional delay value for match (which can be defined by user in our design) goes out as output of the selector unit. The structure of our proposed comparator/selector unit is shown in Fig.\ref{Fig. figures/comperator.JPG}.
	\newline
	
\begin{figure}[h]
		\centerline{\includegraphics[width=0.75\columnwidth]{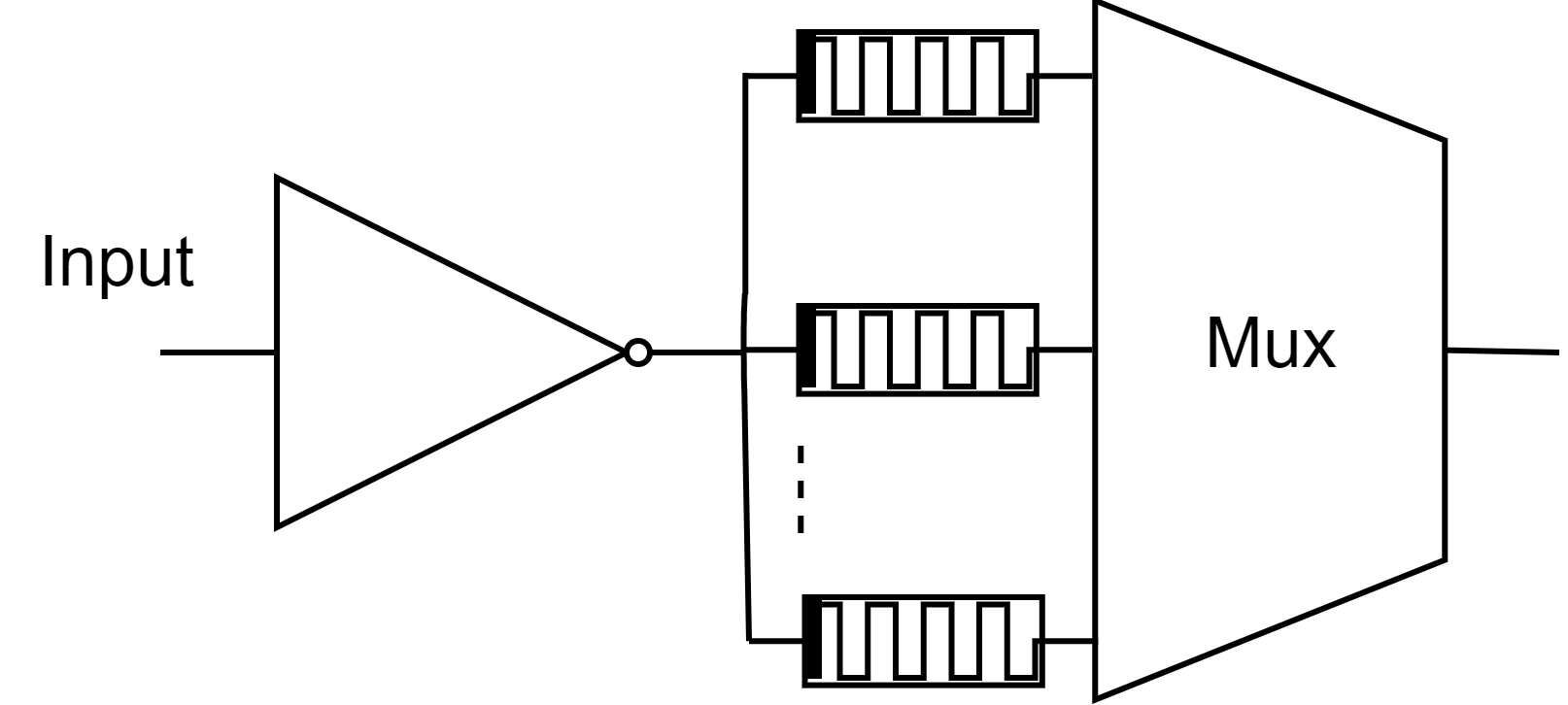}} 
%		\captionsetup{justification=centering}
		\caption{Delay element unit that includes some delay elements to build different delays, and a multiplexer to select the desired delay.}
		\label{Fig. figures/Delay.JPG}
     	\end{figure}

	•-	The delay element (DE):
	\newline
	Delay elements are composed of  :
	\begin{enumerate}
		\item Three input wavefront which are representation of the input signals and are the results of the preceding DEs in the grid
		\item Two corresponding nucleotides as input signals which have to be compared by the element
		\item Three input signals representing the (Match, Mismatch, Gap penalty) values.
		\item One output signal (global OR gate) which represent the output of the Eq. \ref{eq: Min}( $DP_{(i,i)}$)
		\item One output signal (local OR gate) which is designed to perform our desired flexible matrix output and is used for local alignment.
	\end{enumerate}
	
	the propagated output wavefront of each DE is a delay signal with consideration of the corresponding match, mismatch and gap delay penalties. When the other DE’s outputs or signal wavefront reaches an element, a delay is created based on the gap penalty specified for match/mismatch and gap penalty, by propagating the signals through the memristors.
	The other advantage of our design is that it allows the user to program (i.e., dynamically set at runtime) the value of the match, mismatch and gap penalty based on the different applications and give the flexibility to use our approach in cases that merely require re-parameterization of the gap-penalties.
	The structure of our proposed delay element is shown inFig.\ref{Fig. figures/Delay.JPG}. It includes some delay elements to build different delays, and a multiplexer to select the desired delay. As shown in Fig.\ref{Fig. figures/Delay.JPG}, to reduce the area overhead, we have used memristors to implement the delay elements.
	\newline 

\begin{figure}[h]
		\centerline{\includegraphics[width=0.5\columnwidth]{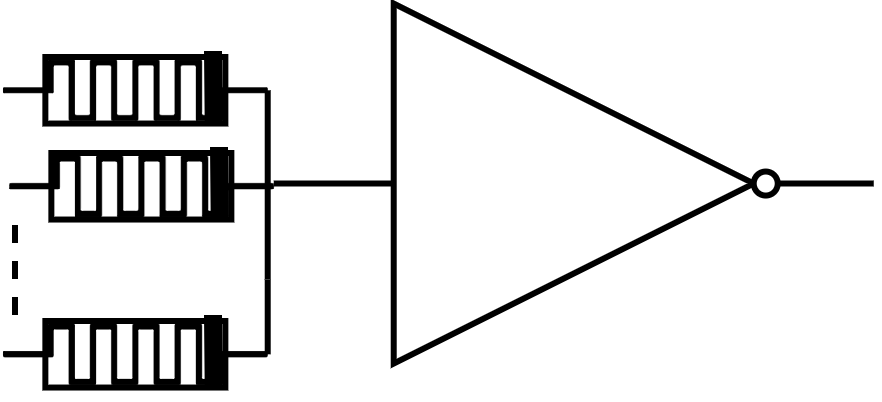}} 
%		\captionsetup{justification=centering}
		\caption{structure of memristor base OR gate in our design.}
		\label{Fig. figures/Or.JPG}
    	\end{figure}

    •-	 Local OR gate:
    \newline
	The local OR gate is used to make it possible to avoid unnecessary latency that is due to variable input length.  OR gate is a proxy for minimization operator, which emerges out the signal that arrives first at the gate. As shown in Fig.\ref{Fig. figures/Or.JPG}, to reduce the area overhead, we have used a memristor-based OR gate for this sake.  
		
	•-	Global OR gate:
	\newline
	The global OR gate is used to implement the minimization operation in the Eq. \ref{eq: Min} . The structure of our proposed global OR gate is shown in Fig.\ref{Fig. figures/Or.JPG}. We have used a memristor-based OR gate for this sake to reduce the area overhead.
	\newline
	•-  The routing network:
	\newline
	Needleman and Wunsch and Smith and Waterman algorithm are well-known dynamic programming algorithm which leads to optimum global and local alignment of a Read against the reference genome.
	In these approaches, a similarity matrix is filled that has to find the local and global alignment score of reads against the corresponding reference sub-string\citep{li2013aligning}. 
	Consider the practical scenario that read data has at most $150_{base pairs}(bp)$ for our comparison. Then we construct our similarity matrix with $131\times131$ dimension based on the BWA-MEM approach. we desire that the processing time of filling the similarity matrix kernel be independent of the read length but because of the fixed similarity matrix dimension, for shorter reads, we incur unnecessary latency.
	
	To avoid this unnecessary latency, we have to contemplate a method which can be flexible with different read length and get output ready from the desired dimensions of similarity matrix. Therefore, we can omit the unnecessary latency which is a reason of not traveling through the entire elements irrespective of its length.

	The original Race Logic design was demonstrated in simulation as an ASIC [13]. Even though this method has advantages in power consumption and substantial improvement in throughput in comparison of the state-of-the-art systolic implementations. But it suffers from the following problems:
	\begin{enumerate}
		\item The original Race Logic design use conventional complementary metal–oxide–semiconductor (CMOS) that has the size, power consumption, read and write time  problems in comparison of our approach.
		\item Traveling through the entire elements irrespective of its length with the fixed similarity matrix dimension design that incur unnecessary latency for shorter read size 
	\end{enumerate}
	
	Our proposed accelerator is runtime-programmable for changing the input data size, which defines the size of the accelerator lattice. For this sake, we have used a nanowire-based routing network which is inspired by FPNI technique \citep{zandevakili2018new}. Field-programmable nanowire interconnect (FPNI) is new hybrid structure with advantages that are mentioned below :
	
	\begin{enumerate}
		\item high flexibilit
		\item low fabrication cost
	\end{enumerate}
	By this technique, we can change the size of the accelerator lattice during the runtime according to the input data size. As shown in Fig.\ref{Fig. figures/Basic-cell.JPG}, our proposed routing network includes some nanowires to access the output of some predefined basic cells, and a selection unit which is controlled by the input data size to select the desired output. Each nanowire is connected through a “via” to the output of local OR gate in the desired basic cell.

	\section{Results}
	
	In this section, the simulation results of the proposed method will be compared with some well-known approaches. Performance of the mentioned methods is evaluated using several criteria such as area, delay and power consumption. In Fig \ref{Fig. figures/F4.png} , the numerical results of the proposed structure for delay parameter are compared with state-of-the-art systolic arrays and Race logic design.
	In general, these are two of the best implementations of the dynamic programming methods that achieve the accuracy and speedup. Therefore, we compare our design to show the consummate performance of our work.

	More details about each of the evaluation criteria will be presented in the following.
	\begin{enumerate}
		\item	Area : To compute the occupied area of the mentioned methods, we have used transistor counting technique in 65nm technology. According to the presented results in Table \ref{area}, the occupied area of the proposed method is compared with two other methods and the results shows that we achieve up to 10fold area improvement.

\begin{table}[]
\addtolength{\tabcolsep}{-3pt}
\begin{center}
\caption{}
\label{area}
\begin{tabular}{|l|lll|lll}
\cline{1-4}
                                                      &                               & Area(nm)                      &            &  &  &  \\ \cline{1-4}
\begin{tabular}[c]{@{}l@{}}Read\\ length\end{tabular} & \multicolumn{1}{l|}{Proposed} & \multicolumn{1}{l|}{Systolic} & Race logic &  &  &  \\ \cline{1-4}
1                                                     & \multicolumn{1}{l|}{8.51E+02} & \multicolumn{1}{l|}{7.34E+04} & 9.18E+03   &  &  &  \\
2                                                     & \multicolumn{1}{l|}{3.40E+03} & \multicolumn{1}{l|}{1.18E+05} & 2.09E+04   &  &  &  \\
4                                                     & \multicolumn{1}{l|}{1.36E+04} & \multicolumn{1}{l|}{2.34E+05} & 7.31E+04   &  &  &  \\ \cline{1-4}

\end{tabular}
\end{center}
\end{table}

		\item	Delay :
		We need an electrical model of the nano wires, junctions and CMOS components to calculate the delay of the proposed structure. For this sake, we have used the electrical model proposed in \citep{snider2007nano} for FPNI structure. The electrical model for a simple circuit is shown in Fig \ref{Fig. figures/F3.PNG}. Some of the model parameters such as closed-junction resistance, the capacitance and resistance per unit length and geometry of the wires are also listed in Table \ref{tab:my-table} \citep{snider2007nano}. In this paper, we have used the HSpice tool to calculate the delay of the proposed structure. According to the presented results in Fig \ref{Fig. figures/F4.png}. %In addition, Fig \ref{Fig. figures/Speedup2.eps} shows our design's speedup in comparison of two other methods. Besides, 
		Fig. \ref{Fig. figures/Fixedlengthspeedup.png} shows how our design flaunt himself in case of fixed length matrix dimension implementation.
		\begin{figure}[h]
			\centerline{\includegraphics[width=\columnwidth]{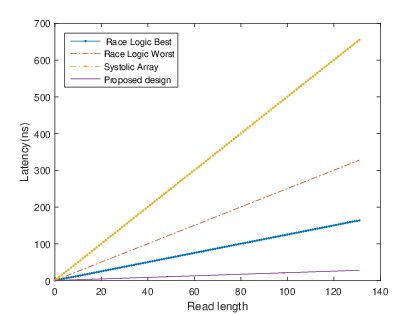}} 
%			\captionsetup{justification=centering}
			\caption{Latency of proposed method in comparison of the state-of-the-art systolic array and race logic designs.}
			\label{Fig. figures/F4.png}
		\end{figure}
		
%		\begin{figure}[h]
%			\centerline{\includegraphics[width=\columnwidth]{figures/Speedup2.eps}} 
%			\captionsetup{justification=centering}
%			\caption{Our proposed design speedup in comparison of Systolic arrays and Race logic design.}
%			\label{Fig. figures/Speedup2.eps}
%		\end{figure}

		\begin{table}[h]
		\addtolength{\tabcolsep}{-3pt}
        \begin{center}
			\caption{Experimental parameters for FPNI architecture \citep{snider2007nano}}
			\label{tab:my-table}
			\begin{tabular}{|l|l|l|}
				\hline
				Parameter & Description                & FPNI 30 nm        \\ \hline
				Pnano     & Nanowire pitch             & 30nm              \\ \hline
				Wnano     & Nanowire width             & 15 nm             \\ \hline
				Wpin      & Pin diameter               & 90 nm             \\ \hline
				Wpinvar   & Pin size variation         & 20 nm             \\ \hline
				Walign    & Alignment error            & 40 nm             \\ \hline
				Wsep      & Pin/wire separation        & 15 nm             \\ \hline
				Rclosed   & Closed junction resistance & 24 K             \\ \hline
				p         & On/off resistance ratio    & \textgreater{}200 \\ \cline{2-3} 
				& Nanowire resistivity       & 8u cm            \\ \cline{2-3} 
				& Nanowire length            & 7115 nm           \\ \cline{2-3} 
				& Nanowire resistance        & 2.53 K           \\ \hline
			\end{tabular}
			\end{center}
		\end{table}
	
		\begin{figure}[h]
			\centerline{\includegraphics[width=\columnwidth]{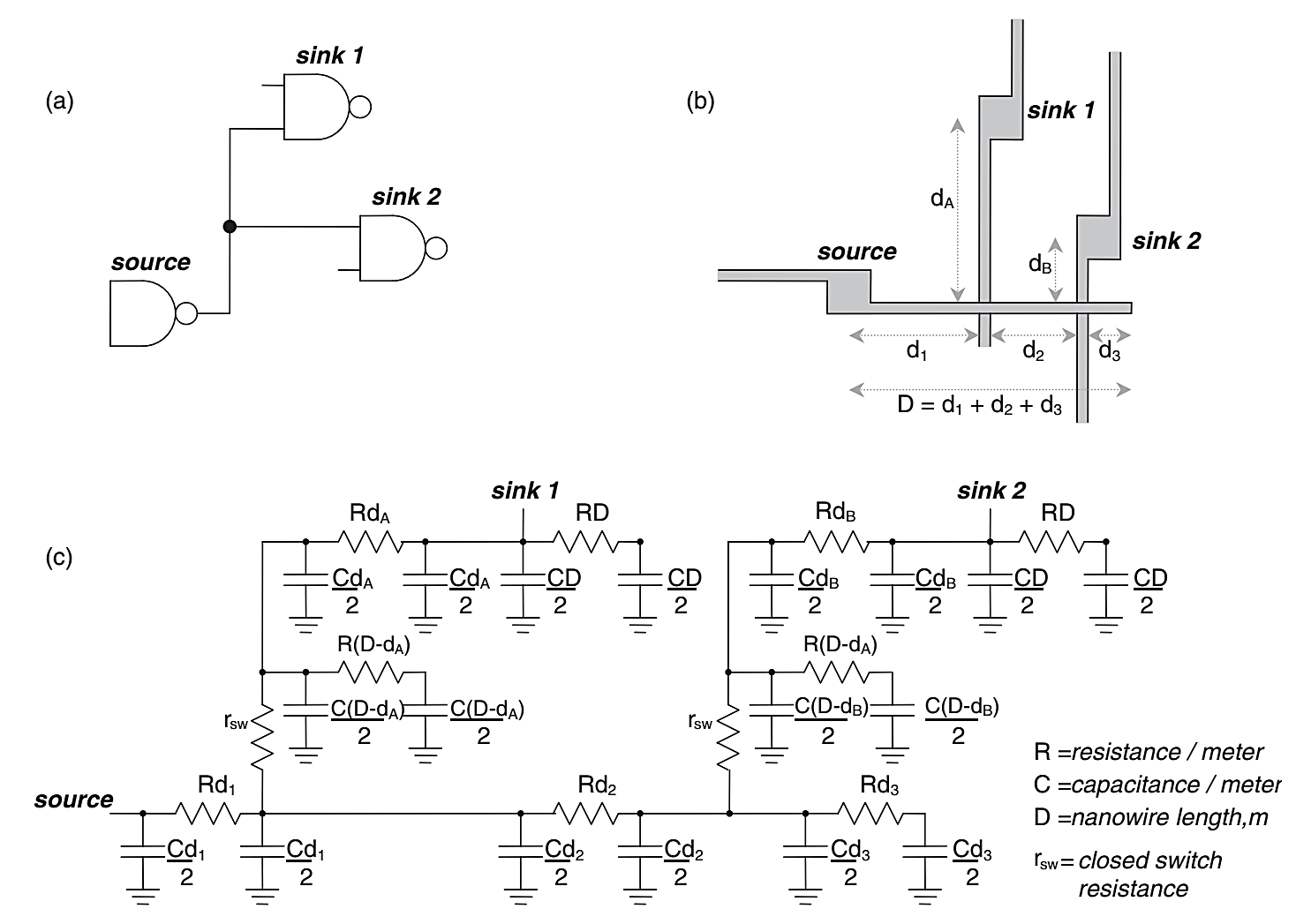}} 
%			\captionsetup{justification=centering}
			\caption{(a) A signal with a fan-out of 2 (b) the implemented form by the nanowires (c) the electrical model \citep{snider2007nano}.}
			\label{Fig. figures/F3.PNG}
		\end{figure}

				\begin{figure}[h]
			\centerline{\includegraphics[width=\columnwidth]{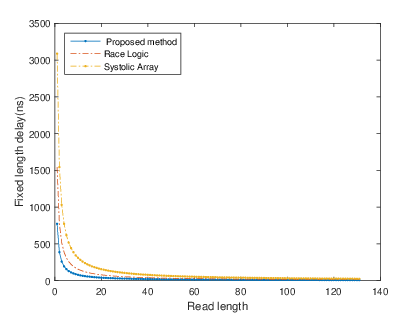}} 
%			\captionsetup{justification=centering}
			\caption{Delay ratio of proposed method, Systolic array and Race logic with considering the fixed $131\times131$ Smith-waterman matrix dimension in different read lengths.}
			\label{Fig. figures/Fixedlengthspeedup.png}
		\end{figure}

		\item	Power consumption :

		Power consumption of the proposed structure is evaluated using the formula proposed in \citep{snider2007nano}:
		
		\begin{equation}
		\label{eq: Power}
		\begin{array}{c} \vspace{8px}
		
		Dynamic-power = \frac{1}{2}ANCV_{dd}^2f
		\end{array}
		\end{equation}

		Where A is the average ‘activity’ of a signal, N is the number of allocated nanowires, C is the capacitance of a single nanowire, $V_{dd}$ is the supply voltage used by the CMOS, and f is the maximum clock frequency determined by timing analysis. To calculate the power consumption of the proposed structure, we have used the HSpice tool. According to the presented results in Fig. \ref{Fig. figures/finalpowercomparison.png}, we compare our design with systolic arrays and Race logic approach and results shows those designs are power hungry in comparison of our memristor-base design.%in case of some matrix dimensions that was able to implement for us.
		
		\begin{figure}[h]
			\centerline{\includegraphics[width=\columnwidth]{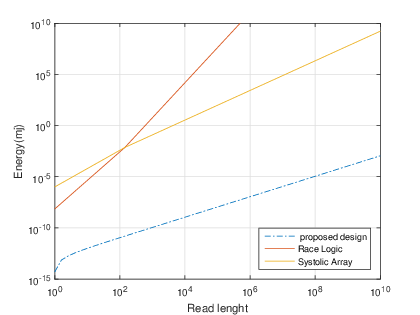}} 
%			\captionsetup{justification=centering}
			\caption{Power consumption of our proposed design in comparison of Systolic arrays and Race logic design.}
			\label{Fig. figures/finalpowercomparison.png}
		\end{figure}

	\end{enumerate}
	\newpage
	\section{Conclusion}
	
	We present a new memristor based smith-waterman matrix implementation that achieves more than 6 times speedup in comparison of the state-of-the-art Race logic approach and 22 times speedup than the systolic arrays implementation. We shows how our design give this flexibility to get the matrix output depends on the different input dimension without suffering from the unnecessary latency.Our implementation achieves up to 600x speedup with considering the fixed $131\times131$ Smith-waterman matrix dimension by testing different read lengths. We also achieved at least 10x improvements in area overhead and also {$10^5$}x improvements in power. Furthermore, our approach can be more and more practical and optimum in case of presenting programmable penalty matches which gives 
    Initiative to change them based on the biological Application.

\clearpage	
\newpage
    \section{References}
	\bibliographystyle{apacite}

	\bibliography{References}
	
\end{document}